\documentclass[12pt]{iopart}

\usepackage{graphicx}
\usepackage{hyperref}
\usepackage{cite}
\usepackage{multirow}
\usepackage{mathrsfs}

\begin{document}

\title{Optimal synthesis of the Fredkin gate in a multilevel system}

\author{Wen-Qiang Liu$^1$ and Hai-Rui Wei$^{1,2}$*}

\address{$^1$School of Mathematics and Physics, University of Science and Technology Beijing, Beijing 100083, China

$^2$Centre for Quantum Technologies, National University of Singapore, 3 Science Drive 2, Singapore 117543}

\ead{*hrwei@ustb.edu.cn}

\begin{abstract}
The optimal cost of a three-qubit Fredkin gate is 5 two-qubit entangling gates, and the overhead climbs to 8 when restricted to controlled-not (CNOT) gates. By harnessing higher-dimensional Hilbert spaces, we reduce the cost of a three-qubit Fredkin gate from 8 CNOTs to 5 nearest-neighbor CNOTs. We also present construction of an $n$-control-qubit Fredkin gate with $2n+3$ CNOTs and $2n$ single-qudit operations. Finally, we design deterministic and nondeterministic three-qubit Fredkin gates in photonic architectures. The cost of a nondeterministic three-qubit Fredkin gate is further reduced to 4 nearest-neighbor CNOTs, and the success of such a gate is heralded by a single-photon detector. Our insights bridge the gap between the theoretical lower bound and the current best result for the $n$-qubit quantum computation.

\end{abstract}


\section{Introduction}\label{sec1}

Quantum computing promises great advantages over its classical counterpart and may be used to solve intractable problems in many areas \cite{book}.
A great challenge in building a full-scale quantum computer is the large number of basic gates required, even in small quantum circuits. The cost of a quantum circuit is usually measured by the number of controlled-NOT (CNOT) gates. Several methods have been used to minimize the number of CNOT gates required in a given circuit, including orthogonal-triangular  \cite{universal}, cosine-sine matrix \cite{CSD}, odd-even \cite{OED}, Khanja and Glaser \cite{KGD}, concurrence canonical \cite{CCD}, and quantum Shannon decompositions (QSD) \cite{QSD}. Unfortunately, a gap continues to remain between the current minimum number of CNOTs determined with QSD ($(23/48)\times 4^n-(3/2)\times 2^n+4/3$) and the unstructured theoretical lower bound ($(4^n-3n-1)/4$) \cite{lower-bound} for an $n$-qubit quantum circuit.  A specific optimal quantum gate usually cannot be produced using the above approaches.

The Fredkin (controlled-swap) gate is a fundamental multi-qubit gate. With the help of Hadamard gates, it can be used to simulate arbitrary multi-qubit quantum computations \cite{Fredkin}. Moreover, a Fredkin gate has been applied to quantum algorithms \cite{algorithm1,algorithm2}, quantum fingerprinting \cite{fingerprinting0},  quantum state preparation \cite{preparation},  quantum state estimation \cite{state-estimation}, optimal cloning \cite{cloning1}, etc. Early in 1995, Chau and Wilczek \cite{Fredkin-six} decomposed a three-qubit Fredkin gate into 6 two-body operators. In 1996, Smolin and DiVincenzo \cite{Fredkin-five1} decomposed a three-qubit Fredkin gate into 5 specific two-qubit entangling gates. In 2015, Yu and Ying \cite{Fredkin-five2} proved theoretically that 5 two-qubit gates are sufficient and necessary for implementing a three-qubit Fredkin gate, but a concrete circuit was not provided. In 2015, Ivanov \emph{et al.} \cite{Ivanov} presented a three-qubit Fredkin gate with 4 globe two-qubit gates or 5 nearest-neighbor interactions. If we further restrict our attention to CNOTs, the overhead of a three-qubit Fredkin gate will increase to 8 CNOTs \cite{Fredkin-eight}, which is less desirable than the minimum of 5 two-qubit gates \cite{Fredkin-five2}. In addition, the simplified synthesis of an $n$-qubit Fredkin gate in terms of CNOTs and single-qubit gates is not presented today.

By transforming the target qubit into a qutrit, Ralph \emph{et al.} \cite{Ralph} and Lanyon \emph{et al.} \cite{Lanyon} reduced the length of a Toffoli gate from 6 CNOTs to 3 CNOTs. By exploiting qudit catalysis, Ionicioiu \emph{et al.} \cite{Ionicioiu} reduced the cost of a generalized Toffoli gate from $O(n^2)$ two-qubit gates to $n$ two-particle gates. With the help of an accessory Hilbert space, Li \emph{et al.} \cite{Li-wen-dong} optimized an $n$-qubit universal quantum circuit with $(5/16)\times 4^n-(5/4)\times2^n+2n$ CNOTs when $n$ is even and $(5/16)\times 4^n-2^n+2(n-1)$ CNOTs when $n$ is odd. Therefore, multi-level physical systems might provide an alternative method for further simplifying quantum circuits. In other words, the cost of a Fredkin gate might be further reduced by using auxiliary dimensions or degrees of freedom (DOFs).

In this paper, we present a procedure for constructing Fredkin gate circuits, including one-control-qubit and $n$-control-qubit Fredkin gates, in terms of CNOTs and single-qudit operations, where the first target qubit in a Fredkin gate is allowed to be a temporary multi-level system during the gate operation. Synthesis of an $n$-control-qubit Fredkin involves only $2n+3$ CNOTs and $2n$ single-qudit operations. Our three-qubit Fredkin beats  the constructions based on 5 two-qubit entangling gates \cite{Fredkin-five1,Fredkin-five2},  6 specific two-body gates \cite{Fredkin-six}, and  8 CNOTs \cite{Fredkin-eight}, in terms of source overheads. Finally, we present deterministic and nondeterministic optical architectures for implementing a three-qubit Fredkin gate, and the gate success is heralded by a single-photon detector. The cost of such a Fredkin gate is further reduced to 4 nearest-neighbor CNOTs.


\section{Deterministic Fredkin gates using multi-level systems} \label{Sec2}


\subsection{Synthesis of three-qubit Fredkin gate using qutrit} \label{Sec2-1}

\begin{figure} [htp] 
\centering
\includegraphics[width=8.5 cm,angle=0]{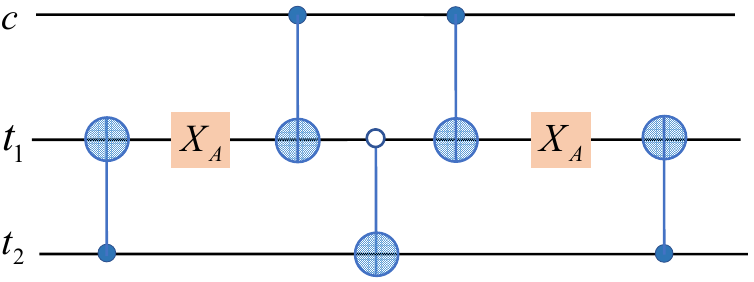}
\caption{Optimal synthesis of a three-qubit Fredkin gate using CNOTs and single-qutrit operations. The CNOT gate acts on the qubit-level $|0\rangle$ and $|1\rangle$ in the normal manner. The $X_A$ operator exchanges the states between $|0\rangle$ and $|2\rangle$. The controls ``$\circ$'' and ``$\bullet$'' are turned on for inputs ``$|0\rangle$'' and ``$|1\rangle$'', respectively.}
\label{Fredkin1}
\end{figure}

As shown in Fig. \ref{Fredkin1}, optimal synthesis of our three-qubit Fredkin gate involves only 5 nearest-neighbor CNOTs and 2 single-qutrit operations. Optimization is achieved by expanding the first target to a qutrit (labeled $|0\rangle$, $|1\rangle$, and $|2\rangle$); others are labeled as common logic states $|0\rangle$ and $|1\rangle$ (i.e., qubit). All CNOTs act on the qubit-level in the usual manner.

We describe our synthesis in some detail. Suppose a three-qubit state is initially prepared as follows:
\begin{eqnarray}              \label{eq1}
\fl\qquad\quad|\psi_0\rangle=\alpha_1|0\rangle_c|0\rangle_{t_1}|0\rangle_{t_2}+\alpha_2|0\rangle_c|0\rangle_{t_1}|1\rangle_{t_2}
+\alpha_3|0\rangle_c|1\rangle_{t_1}|0\rangle_{t_2}+\alpha_4|0\rangle_c|1\rangle_{t_1}|1\rangle_{t_2} \nonumber\\
\fl\qquad\qquad\quad+\,\alpha_5|1\rangle_c|0\rangle_{t_1}|0\rangle_{t_2}+\alpha_6|1\rangle_c|0\rangle_{t_1}|1\rangle_{t_2}
+\alpha_7|1\rangle_c|1\rangle_{t_1}|0\rangle_{t_2}+\alpha_8|1\rangle_c|1\rangle_{t_1}|1\rangle_{t_2}.
\end{eqnarray}
Here, the coefficients $\alpha_i$ $(i=1,2,\cdots,8)$ are arbitrary complex numbers satisfying the normalization condition $\sum_{i=1}^8 |\alpha_i|^2=1$. The subscripts $c$, $t_1$, and $t_2$ represent the control qubit $c$, first target qubit $t_1$, and second target qubit $t_2$, respectively.

First, as shown in Fig. \ref{Fredkin1}, a CNOT gate with $t_2$ as the control qubit and $t_1$ as the target qubit is used to obtain the following:
\begin{eqnarray}              \label{eq2}
\fl\qquad\quad|\psi_1\rangle=\alpha_1|0\rangle_c|0\rangle_{t_1}|0\rangle_{t_2}+\alpha_2|0\rangle_c|1\rangle_{t_1}|1\rangle_{t_2}
+\alpha_3|0\rangle_c|1\rangle_{t_1}|0\rangle_{t_2}+\alpha_4|0\rangle_c|0\rangle_{t_1}|1\rangle_{t_2} \nonumber\\
\fl\qquad\quad\qquad+\,\alpha_5|1\rangle_c|0\rangle_{t_1}|0\rangle_{t_2}+\alpha_6|1\rangle_c|1\rangle_{t_1}|1\rangle_{t_2}
+\alpha_7|1\rangle_c|1\rangle_{t_1}|0\rangle_{t_2}+\alpha_8|1\rangle_c|0\rangle_{t_1}|1\rangle_{t_2}.
\end{eqnarray}

Second, the operator $X_A$ is defined in the following matrix representation:
\begin{eqnarray}                      \label{eq3}
\qquad\qquad\qquad\quad U_{X_A}=\left(\begin{array}{ccc}
0 & 0 & 1  \\
0 & 1 & 0 \\
1 & 0 & 0
\end{array}\right)
\end{eqnarray}
The basis $\{|0\rangle, |1\rangle, |2\rangle\}$ is used to move information from the $|0\rangle$ state of $t_1$ to the $|2\rangle$ state for bypassing the three subsequent CNOTs. That is, the three subsequent CNOTs only operate on the subspaces $|0\rangle_c|1\rangle_{t_1}|1\rangle_{t_2}$,
$|0\rangle_c|1\rangle_{t_1}|0\rangle_{t_2}$,
$|1\rangle_c|1\rangle_{t_1}|1\rangle_{t_2}$, and
$|1\rangle_c|1\rangle_{t_1}|0\rangle_{t_2}$.

Third, the CNOT gate with $c$ ($t_1$) as the control (target) qubit, $\overline{\rm CNOT}$ gate with $t_1$ ($t_2$) as the control (target) qubit, and CNOT gate with $c$ ($t_1$) as the control (target) qubit are preformed in succession. This arrangement of three gates transforms the state of the whole system from
\begin{eqnarray}              \label{eq4}
\fl\qquad\quad|\psi_2\rangle=\alpha_1|0\rangle_c|2\rangle_{t_1}|0\rangle_{t_2}+\alpha_2|0\rangle_c|1\rangle_{t_1}|1\rangle_{t_2}
+\alpha_3|0\rangle_c|1\rangle_{t_1}|0\rangle_{t_2}+\alpha_4|0\rangle_c|2\rangle_{t_1}|1\rangle_{t_2} \nonumber\\
\fl\qquad\quad\qquad+\,\alpha_5|1\rangle_c|2\rangle_{t_1}|0\rangle_{t_2}+\alpha_6|1\rangle_c|1\rangle_{t_1}|1\rangle_{t_2}
+\alpha_7|1\rangle_c|1\rangle_{t_1}|0\rangle_{t_2}+\alpha_8|1\rangle_c|2\rangle_{t_1}|1\rangle_{t_2}
\end{eqnarray}
to
\begin{eqnarray}              \label{eq8}
\fl\qquad\quad|\psi_3\rangle=
\alpha_1|0\rangle_c|2\rangle_{t_1}|0\rangle_{t_2}+
\alpha_2|0\rangle_c|1\rangle_{t_1}|1\rangle_{t_2}+
\alpha_3|0\rangle_c|1\rangle_{t_1}|0\rangle_{t_2}+
\alpha_4|0\rangle_c|2\rangle_{t_1}|1\rangle_{t_2}\nonumber\\
\fl\qquad\quad\qquad+\,\alpha_5|1\rangle_c|2\rangle_{t_1}|0\rangle_{t_2}+
\alpha_6|1\rangle_c|1\rangle_{t_1}|0\rangle_{t_2}+
\alpha_7|1\rangle_c|1\rangle_{t_1}|1\rangle_{t_2}+
\alpha_8|1\rangle_c|2\rangle_{t_1}|1\rangle_{t_2}.
\end{eqnarray}
Here, the $\overline{\rm CNOT}$ gate flips the state of the target qubit if and only if (iff) the control qubit is in the state $|0\rangle$. Hence, the $\overline{\rm CNOT}$ and CNOT gates are equivalent up to two local bit-flip operations $\sigma_x=|0\rangle\langle1|+|1\rangle|\langle0|$, i.e.,
\begin{eqnarray}              \label{eq7}
\qquad\qquad\quad{\rm CNOT}=\sigma_x \otimes I_2 \cdot \overline{\rm CNOT}\cdot \sigma_x \otimes I_2 .
\end{eqnarray}

Finally, $X_A$ is used to contract $t_1$ into the original two-dimensional space, and the CNOT gate with $t_2$ ($t_1$) as the control (target) qubit is used again to obtain the state
\begin{eqnarray}              \label{eq10}
\fl\qquad\quad|\psi_4\rangle=
\alpha_1|0\rangle_c|0\rangle_{t_1}|0\rangle_{t_2}+
\alpha_2|0\rangle_c|0\rangle_{t_1}|1\rangle_{t_2}+
\alpha_3|0\rangle_c|1\rangle_{t_1}|0\rangle_{t_2}+
\alpha_4|0\rangle_c|1\rangle_{t_1}|1\rangle_{t_2}\nonumber\\
\fl\qquad\quad\qquad+\,
\alpha_5|1\rangle_c|0\rangle_{t_1}|0\rangle_{t_2}+
\alpha_6|1\rangle_c|1\rangle_{t_1}|0\rangle_{t_2}+
\alpha_7|1\rangle_c|0\rangle_{t_1}|1\rangle_{t_2}+
\alpha_8|1\rangle_c|1\rangle_{t_1}|1\rangle_{t_2}.
\end{eqnarray}

From Eqs. (\ref{eq1})--(\ref{eq10}), one can see that a three-qubit Fredkin gate can be synthesized from 5 nearest-neighbor CNOTs and 2 single-qutrit gates (see Fig. \ref{Fredkin1}). Our synthesis is optimal as the CNOT-count test suggests its optimization in Ref. \cite{Fredkin-five2}.

\subsection{Synthesis of an $n$-control-qubit Fredkin gate using qudit} \label{Sec2-2}

\begin{figure} [htp] 
\centering
\includegraphics[width=9.3 cm,angle=0]{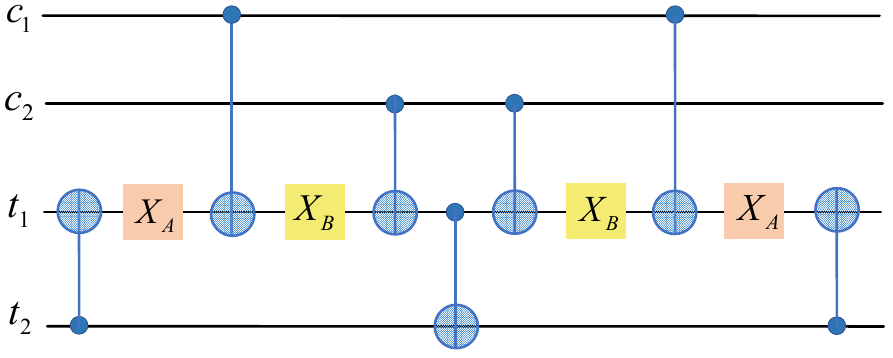}
\caption{Synthesis of a two-control-qubit Fredkin gate using a four-dimensional Hilbert space. Operations $X_A$ and $X_B$ complete the transformations $|0\rangle \leftrightarrow |2\rangle$ and $|1\rangle \leftrightarrow |3\rangle$, respectively.} \label{CCSWAP1}
\end{figure}

Our method can be generalized to simulate an $n$-control-qubit Fredkin gate by expanding the first target qubit to $(n+2)$ levels. The $n$-control-qubit Fredkin gate exchanges information with the two target qubits iff the $n$-control qubits are all in the $|1\rangle$ state. Fig. \ref{CCSWAP1} specifically describes the synthesis of a two-control-qubit Fredkin using 7 CNOTs and 4 single-qudit gates. The increased efficiency requires use of $X_A$ ($X_B$) to exchange quantum information between $|0\rangle$ ($|1\rangle$) and $|2\rangle$ ($|3\rangle$), thus subsequent operations can be bypassed.

\begin{figure} [htp] 
\centering
\includegraphics[width=12 cm,angle=0]{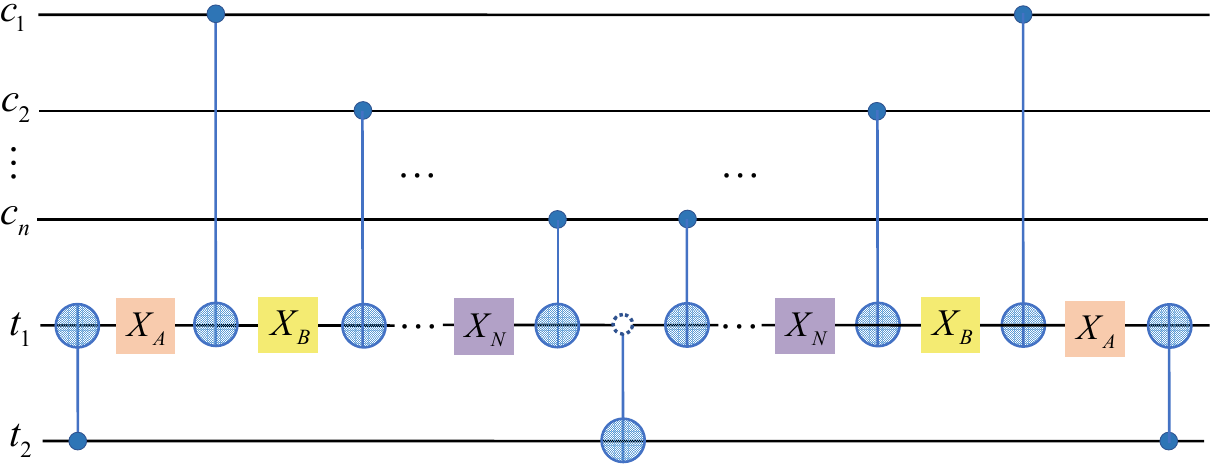}
\caption{Synthesis of an $n$-control-qubit Fredkin. Operations $X_A$, $X_B$, $\cdots$, $X_N$ complete the transformations $|0\rangle \leftrightarrow |2\rangle$, $|1\rangle \leftrightarrow |3\rangle$, $\cdots$, $|0\rangle \leftrightarrow |n+1\rangle$ if $n$ is odd, or $|0\rangle \leftrightarrow |2\rangle$, $|1\rangle \leftrightarrow |3\rangle$, $\cdots$, $|1\rangle \leftrightarrow |n+1\rangle$ if $n$ is even. The control node of the middle CNOT is turned on ``$\circ$'' (corresponding to $|0\rangle$) when $n$ is odd and turned on  ``$\bullet$'' (corresponding to $|1\rangle$) when $n$ is even, respectively. } \label{n-control}
\end{figure}

Generally, as shown in Fig. \ref{n-control}, $2n+3$ CNOTs and $2n$ single-qudit gates are sufficient for constructing an $n$-control-qubit Fredkin gate by allowing the first target qubit to temporarily take $(n+2)$ levels. Evidently, the polynomial number of CNOTs in our scheme is far less than the minimum required number of CNOTs $O(n^2)$  \cite{universal}, and a growing advantage of our presented scheme emerges as $n$ increases. Our simplified outcomes indicate that the gap between the theoretical lower bound \cite{lower-bound} and the current minimum result might be bridged  by harnessing higher-dimensional Hilbert spaces.

\section{ Photonic architecture of the Fredkin gate} \label{Sec3}

Multi-level systems are necessary for the technique we use to simplify a Fredkin gate. Fortunately, photons serve as outstanding candidates for encoding quantum information and naturally offer multi-level structures owing to their wide range of accessible DOFs, including polarization, spatial-mode, time-bin, frequency, and orbital momentum.
A linear polarization CNOT gate with 0.75 success probability was prepared by Knill, Laflamme, and Milburn \cite{KLM} in 2001.
A polarization CNOT gate with 0.25 success probability can be obtained when using an entangled photon pair as resources  \cite{Pittman}. A measurement-based optical CNOT gate was demonstrated \cite{one-way2} in 2007. Moreover, parallel and hyperparallel deterministic optical CNOT gates prepared from photon-matter emitters have been proposed in recent years \cite{CNOT2,CNOT3,hyper-CNOT2,hyper-CNOT3} and cross-Kerr approaches  \cite{Kerr-CNOT}.


\subsection{Deterministic three-qubit optical Fredkin gate} \label{Sec3-1}

\begin{figure} [!h]
\centering
\includegraphics[width=8.3 cm,angle=0]{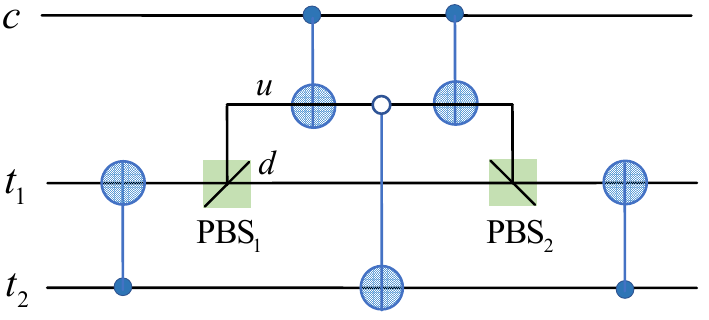}
\caption{Optical implementation of the deterministic Fredkin gate. Polarizing beam splitters (PBSs) transmit the horizontal polarization $H$ and reflect the vertical polarization $V$. } \label{deter-opti-Fredkin}
\end{figure}

Fig. \ref{deter-opti-Fredkin} shows the scheme we designed for implementing three-qubit optical Fredkin gate using a qutrit. The gate qubits are encoded in the polarization DOF of a single photon such that $|H\rangle\equiv|0\rangle$ and $|V\rangle\equiv|1\rangle$. Here, $|H\rangle$ ($|V\rangle$) represents a horizontally (vertically) polarized photon. $X_A$ is a key ingredient in our program, which is achieved through the use of two polarizing beam splitters (PBS$_1$ and PBS$_2$). PBS$_1$ and PBS$_2$ transmit the $H$-polarized component and reflect the $V$-polarized component, respectively. The polarization-encoded optical CNOT gate can be fabricated to apply to our schemes by assisting an entangled photon source  \cite{Pittman}.


The first CNOT gate with $t_2$ ($t_1$) as the control (target) qubit transforms the total state of the composite system from the initial state  $|\varphi_0\rangle$  into $|\varphi_1\rangle$. Here,
\begin{eqnarray}              \label{eq11}
\fl|\varphi_0\rangle=\alpha_1|H\rangle_c|H\rangle_{t_1}|H\rangle_{t_2}+\alpha_2|H\rangle_c|H\rangle_{t_1}|V\rangle_{t_2}
+\alpha_3|H\rangle_c|V\rangle_{t_1}|H\rangle_{t_2}+\alpha_4|H\rangle_c|V\rangle_{t_1}|V\rangle_{t_2}\nonumber\\
\fl\quad\quad+\,\alpha_5|V\rangle_c|H\rangle_{t_1}|H\rangle_{t_2}+\alpha_6|V\rangle_c|H\rangle_{t_1}|V\rangle_{t_2}
+\alpha_7|V\rangle_c|V\rangle_{t_1}|H\rangle_{t_2}+\alpha_8|V\rangle_c|V\rangle_{t_1}|V\rangle_{t_2},
\end{eqnarray}
\begin{eqnarray}              \label{eq12}
\fl|\varphi_1\rangle=\alpha_1|H\rangle_c|H\rangle_{t_1}|H\rangle_{t_2}+\alpha_2|H\rangle_c|V\rangle_{t_1}|V\rangle_{t_2}
+\alpha_3|H\rangle_c|V\rangle_{t_1}|H\rangle_{t_2}+\alpha_4|H\rangle_c|H\rangle_{t_1}|V\rangle_{t_2} \nonumber\\
\fl\quad\quad+\,\alpha_5|V\rangle_c|H\rangle_{t_1}|H\rangle_{t_2}+\alpha_6|V\rangle_c|V\rangle_{t_1}|V\rangle_{t_2}
+\alpha_7|V\rangle_c|V\rangle_{t_1}|H\rangle_{t_2}+\alpha_8|V\rangle_c|H\rangle_{t_1}|V\rangle_{t_2}.
\end{eqnarray}

Subsequently, PBS$_1$ respectively transforms
$|H\rangle_c|V\rangle_{t_1}|V\rangle_{t_2}$,
$|H\rangle_c|V\rangle_{t_1}|H\rangle_{t_2}$,
$|V\rangle_c|V\rangle_{t_1}|V\rangle_{t_2}$, and
$|V\rangle_c|V\rangle_{t_1}|H\rangle_{t_2}$
into
$|H\rangle_c|V,u\rangle_{t_1}|V\rangle_{t_2}$,
$|H\rangle_c|V,u\rangle_{t_1}|H\rangle_{t_2}$,
$|V\rangle_c|V,u\rangle_{t_1}|V\rangle_{t_2}$, and
$|V\rangle_c|V,u\rangle_{t_1}|H\rangle_{t_2}$
for interacting with the subsequent three CNOTs.
Meanwhile, it respectively transforms
$|H\rangle_c|H\rangle_{t_1}|H\rangle_{t_2}$,
$|H\rangle_c|H\rangle_{t_1}|V\rangle_{t_2}$,
$|V\rangle_c|H\rangle_{t_1}|H\rangle_{t_2}$, and
$|V\rangle_c|H\rangle_{t_1}|V\rangle_{t_2}$
into
$|H\rangle_c|H,d\rangle_{t_1}|H\rangle_{t_2}$,
$|H\rangle_c|H,d\rangle_{t_1}|V\rangle_{t_2}$,
$|V\rangle_c|H,d\rangle_{t_1}|H\rangle_{t_2}$, and
$|V\rangle_c|H,d\rangle_{t_1}|V\rangle_{t_2}$
to bypass the subsequent three CNOTs. Here, $u$ and $d$ are two spatial modes of the first target photon $t_1$.
Therefore, after PBS$_1$ and the subsequent three CNOTs are used in succession, the state of the system becomes
\begin{eqnarray}              \label{eq13}
\fl\qquad\qquad|\varphi_2\rangle=
\alpha_1|H\rangle_c|H,d\rangle_{t_1}|H\rangle_{t_2}+
\alpha_2|H\rangle_c|V,u\rangle_{t_1}|V\rangle_{t_2}+
\alpha_3|H\rangle_c|V,u\rangle_{t_1}|H\rangle_{t_2}\nonumber\\
\fl\qquad\qquad\quad\;\;\;\,+\,\alpha_4|H\rangle_c|H,d\rangle_{t_1}|V\rangle_{t_2}+
\alpha_5|V\rangle_c|H,d\rangle_{t_1}|H\rangle_{t_2}+
\alpha_6|V\rangle_c|V,u\rangle_{t_1}|H\rangle_{t_2}\nonumber\\
\fl\qquad\qquad\quad\;\;\;\,+\, \alpha_7|V\rangle_c|V,u\rangle_{t_1}|V\rangle_{t_2}+
\alpha_8|V\rangle_c|H,d\rangle_{t_1}|V\rangle_{t_2}.
\end{eqnarray}

Third, PBS$_2$ contracts the logical states $|H,u\rangle_{t_1}$, $|V,u\rangle_{t_1}$ $|H,d\rangle_{t_1}$, and $|V,d\rangle_{t_1}$ into the original polarized states $|H\rangle_{t_1}$ and $|V\rangle_{t_1}$ . Therefore, PBS$_2$ and the last CNOT gate transform $|\varphi_2\rangle$ into
\begin{eqnarray}              \label{eq15}
\fl|\varphi_3\rangle=\alpha_1|H\rangle_c|H\rangle_{t_1}|H\rangle_{t_2}+\alpha_2|H\rangle_c|H\rangle_{t_1}|V\rangle_{t_2}
+\alpha_3|H\rangle_c|V\rangle_{t_1}|H\rangle_{t_2}+\alpha_4|H\rangle_c|V\rangle_{t_1}|V\rangle_{t_2} \nonumber\\
\fl\qquad+\,\alpha_5|V\rangle_c|H\rangle_{t_1}|H\rangle_{t_2}+\alpha_6|V\rangle_c|V\rangle_{t_1}|H\rangle_{t_2}
+\alpha_7|V\rangle_c|H\rangle_{t_1}|V\rangle_{t_2}+\alpha_8|V\rangle_c|V\rangle_{t_1}|V\rangle_{t_2}.
\end{eqnarray}

From the aforementioned, one finds that a deterministic optical three-qubit Fredkin gate can, in principle, be prepared with the scheme shown in Fig. \ref{deter-opti-Fredkin}.

\subsection{Heralded three-qubit optical Fredkin gate} \label{Sec3-2}

\begin{figure} [!h]
\centering
\includegraphics[width=8.3 cm,angle=0]{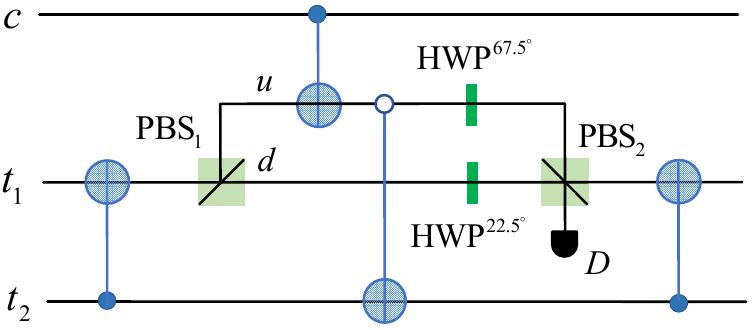}
\caption{Optical implementation of non-deterministic and heralded three-qubit Fredkin gates. The half-wave plate set to 67.5$^\circ$ (HWP$^{67.5^\circ}$) results in $|H\rangle \rightarrow (-|H\rangle+|V\rangle)/\sqrt{2}$ and $|V\rangle \rightarrow (|H\rangle+|V\rangle)/\sqrt{2}$. HWP$^{22.5^\circ}$ completes $|H\rangle \rightarrow (|H\rangle+|V\rangle)/\sqrt{2}$ and $|V\rangle \rightarrow(|H\rangle-|V\rangle)/\sqrt{2}$. $D$ is a single-photon detector.} \label{proba-opti-Fredkin}
\end{figure}

The cost of the above deterministic optical three-qubit Fredkin gate is 5 nearest-neighbor CNOTs, although this gate can be reduced to a probabilistic three-qubit Fredkin gate containing 4 CNOTs. The operation of this gate can be demonstrated with a single-photon detector (see  Fig. \ref{proba-opti-Fredkin}).

First, the same arguments for the deterministic three-qubit optical Fredkin gate show that, the first three CNOTs, HWP$^{67.5^\circ}$ and HWP$^{22.5^\circ}$ transform the joint state of the system to
\begin{eqnarray}              \label{eq19}
\fl\qquad\qquad|\phi_1\rangle=\frac{1}{\sqrt{2}}[\alpha_1|H\rangle_c(|H,d\rangle+|V,d\rangle)_{t_1}|H\rangle_{t_2}+
\alpha_2|H\rangle_c(|H,u\rangle+|V,u\rangle)_{t_1}|V\rangle_{t_2} \nonumber\\
+\,\alpha_3|H\rangle_c(|H,u\rangle+|V,u\rangle)_{t_1}|H\rangle_{t_2}
+\alpha_4|H\rangle_c(|H,d\rangle+|V,d\rangle)_{t_1}|V\rangle_{t_2} \nonumber\\
+\,\alpha_5|V\rangle_c(|H,d\rangle+|V,d\rangle)_{t_1}|H\rangle_{t_2}
+\alpha_6|V\rangle_c(-|H,u\rangle+|V,u\rangle)_{t_1}|H\rangle_{t_2} \nonumber\\
+\,\alpha_7|V\rangle_c(-|H,u\rangle+|V,u\rangle)_{t_1}|V\rangle_{t_2}
+\alpha_8|V\rangle_c(|H,d\rangle+|V,d\rangle)_{t_1}|V\rangle_{t_2}].
\end{eqnarray}
Here, half-wave plates HWP$^{67.5^\circ}$ and HWP$^{22.5^\circ}$ are oriented at 67.5$^\circ$ and $22.5^\circ$ to complete the following transformations:
\begin{eqnarray}                      \label{eq17}
\fl\qquad\;\;{\rm HWP^{67.5^\circ}}|H\rangle &= \frac{1}{\sqrt{2}}(-|H\rangle+|V\rangle), \qquad
{\rm HWP^{67.5^\circ}}|V\rangle &= \frac{1}{\sqrt{2}}(|H\rangle+|V\rangle),\nonumber\\
\fl\qquad\;\;{\rm HWP^{22.5^\circ}}|H\rangle &= \frac{1}{\sqrt{2}}(|H\rangle+|V\rangle), \qquad\;\;\,
{\rm HWP^{22.5^\circ}}|V\rangle &= \frac{1}{\sqrt{2}}(|H\rangle-|V\rangle).
\end{eqnarray}

Subsequently, photons emitted from spatial modes $u$ and $d$ converge at PBS$_2$, and PBS$_2$ transforms the state $|\phi_1\rangle$ to
\begin{eqnarray}              \label{eq20}
\fl|\phi_2\rangle=\frac{1}{\sqrt{2}}(
\alpha_1|H\rangle_c|H\rangle_{t_1}|H\rangle_{t_2}+
\alpha_2|H\rangle_c|V\rangle_{t_1}|V\rangle_{t_2}+
\alpha_3|H\rangle_c|V\rangle_{t_1}|H\rangle_{t_2}+
\alpha_4|H\rangle_c|H\rangle_{t_1}|V\rangle_{t_2}\nonumber\\
\fl\quad\;\;\;\,+\,\alpha_5|V\rangle_c|H\rangle_{t_1}|H\rangle_{t_2}+
\alpha_6|V\rangle_c|V\rangle_{t_1}|H\rangle_{t_2}+
\alpha_7|V\rangle_c|V\rangle_{t_1}|V\rangle_{t_2}\nonumber\\
\fl\quad\;\;\;\,+\,\alpha_8|V\rangle_c|H\rangle_{t_1}|V\rangle_{t_2})+
\frac{1}{\sqrt{2}}(
\alpha_1|H\rangle_c|V,D\rangle_{t_1}|H\rangle_{t_2}
+\alpha_2|H\rangle_c|H,D\rangle_{t_1}|V\rangle_{t_2}\nonumber\\
\fl\quad\;\;\;\,+\,\alpha_3|H\rangle_c|H,D\rangle_{t_1}|H\rangle_{t_2}+
\alpha_4|H\rangle_c|V,D\rangle_{t_1}|V\rangle_{t_2}
+\alpha_5|V\rangle_c|V,D\rangle_{t_1}|H\rangle_{t_2}\nonumber\\
\fl\quad\;\;\;\,-\,\alpha_6|V\rangle_c|H,D\rangle_{t_1}|H\rangle_{t_2}-
\alpha_7|V\rangle_c|H,D\rangle_{t_1}|V\rangle_{t_2}+
\alpha_8|V\rangle_c|V,D\rangle_{t_1}|V\rangle_{t_2}).
\end{eqnarray}
Here, $|H,D\rangle$ and $|V,D\rangle$ denote $H$-polarized and $V$-polarized photons will be detected by a single-photon detector.

Third, the last CNOT gate with $t_2$ ($t_1$) as the control (target) qubit is executed, resulting in
\begin{eqnarray}              \label{eq21}
\fl|\phi_3\rangle=\frac{1}{\sqrt{2}}(
\alpha_1|H\rangle_c|H\rangle_{t_1}|H\rangle_{t_2}+
\alpha_2|H\rangle_c|H\rangle_{t_1}|V\rangle_{t_2}+
\alpha_3|H\rangle_c|V\rangle_{t_1}|H\rangle_{t_2}+
\alpha_4|H\rangle_c|V\rangle_{t_1}|V\rangle_{t_2}\nonumber\\
\fl\quad\;\;\;\,+\,
\alpha_5|V\rangle_c|H\rangle_{t_1}|H\rangle_{t_2}+
\alpha_6|V\rangle_c|V\rangle_{t_1}|H\rangle_{t_2}+
\alpha_7|V\rangle_c|H\rangle_{t_1}|V\rangle_{t_2}\nonumber\\
\fl\quad\;\;\;\,+\,
\alpha_8|V\rangle_c|V\rangle_{t_1}|V\rangle_{t_2})+
\frac{1}{\sqrt{2}}(
\alpha_1|H\rangle_c|V,D\rangle_{t_1}|H\rangle_{t_2}+
\alpha_2|H\rangle_c|H,D\rangle_{t_1}|V\rangle_{t_2}\nonumber\\
\fl\quad\;\;\;\,+\,\alpha_3|H\rangle_c|H,D\rangle_{t_1}|H\rangle_{t_2}+
\alpha_4|H\rangle_c|V,D\rangle_{t_1}|V\rangle_{t_2} +
\alpha_5|V\rangle_c|V,D\rangle_{t_1}|H\rangle_{t_2}\nonumber\\
\fl\quad\;\;\;\,-\,\alpha_6|V\rangle_c|H,D\rangle_{t_1}|H\rangle_{t_2}-
\alpha_7|V\rangle_c|H,D\rangle_{t_1}|V\rangle_{t_2}+
\alpha_8|V\rangle_c|V,D\rangle_{t_1}|V\rangle_{t_2}).
\end{eqnarray}
From Eq. (\ref{eq21}), one can see that if the single-photon detector is activated, the entire system collapses into the following unwanted state:
\begin{eqnarray}              \label{eq22}
\fl|\phi_4\rangle=\frac{1}{\sqrt{2}}(
\alpha_1|H\rangle_c|V\rangle_{t_1}|H\rangle_{t_2}+
\alpha_2|H\rangle_c|H\rangle_{t_1}|V\rangle_{t_2}+
\alpha_3|H\rangle_c|H\rangle_{t_1}|H\rangle_{t_2}+
\alpha_4|H\rangle_c|V\rangle_{t_1}|V\rangle_{t_2} \nonumber\\
\fl\quad\;\;\;\,+\,\alpha_5|V\rangle_c|V\rangle_{t_1}|H\rangle_{t_2}-
\alpha_6|V\rangle_c|H\rangle_{t_1}|H\rangle_{t_2}-
\alpha_7|V\rangle_c|H\rangle_{t_1}|V\rangle_{t_2}+
\alpha_8|V\rangle_c|V\rangle_{t_1}|V\rangle_{t_2}).
\end{eqnarray}
Otherwise, the entire system will collapse into the following desired state:
\begin{eqnarray}              \label{eq23}
\fl|\phi_4'\rangle=\frac{1}{\sqrt{2}}(
\alpha_1|H\rangle_c|H\rangle_{t_1}|H\rangle_{t_2}+
\alpha_2|H\rangle_c|H\rangle_{t_1}|V\rangle_{t_2}+
\alpha_3|H\rangle_c|V\rangle_{t_1}|H\rangle_{t_2}+
\alpha_4|H\rangle_c|V\rangle_{t_1}|V\rangle_{t_2}\nonumber\\
\fl\quad\;\;\;\,+\,\alpha_5|V\rangle_c|H\rangle_{t_1}|H\rangle_{t_2}+
\alpha_6|V\rangle_c|V\rangle_{t_1}|H\rangle_{t_2}+
\alpha_7|V\rangle_c|H\rangle_{t_1}|V\rangle_{t_2}+
\alpha_8|V\rangle_c|V\rangle_{t_1}|V\rangle_{t_2}).
\end{eqnarray}
Therefore, the quantum circuit shown in Fig. \ref{proba-opti-Fredkin} can function as a non-deterministic three-qubit optical quantum Fredkin gate, and the success of such a heralded gate can be demonstrated with a single-photon detector.

\section{Discussion and conclusion} \label{Sec4}

Quantum computation has received increased attentions. Seeking a minimum use of CNOTs is at the core of quantum computing \cite{large-circuit1,large-circuit2}. By harnessing a qudit of the first target information carrier, we illustrated a procedure for simulating an $n$-control-qubit Fredkin gate. Our method helped bridge the gap between the current optimal result and the theoretical lower bound  $(4^n-3n-1)/4$ CNOTs for $n$-qubit quantum circuits. The number of CNOTs implied that our synthesis of a one-control-qubit Fredkin gate is optimal \cite{Fredkin-five2} and the 5 CNOTs are all nearest neighbors. Furthermore, one should note that a nonlocal two-qubit gate is not allowed in general. A long-range CNOT gate between the $1^{\rm st}$ and $3^{\rm rd}$  qubits can be stimulated by 4 nearest-neighbor CNOTs \cite{nearest-neighbor-CNOT}. For the longer-range CNOT gate between the $1^{\rm st}$ qubit and the $4^{\rm th}$ qubit, the number of the nearest-neighbor CNOTs will increase to 8.

Assisted by a further spatial DOF of a single photon, we also designed two compact schemes for implementing a three-qubit deterministic Fredkin gate and a heralded Fredkin gate. The cost of the non-deterministic Fredkin gate can be further reduced to 4 nearest-neighbor CNOTs, which is superior to the post-selected \cite{Gong-Guo}, partial-SWAP-based \cite{Fiurasek}, and cross-Kerr-based constructions \cite{Fredkin-cross}.
Indeed, the superconducting circuit, diamond nitrogen-vacancy (NV) defect center, and optical system can provide multiple levels to implement universal quantum gates. The two computing states and one auxiliary state can be encoded in the higher energy level states in superconducting circuits \cite{3-level}. In superconducting circuits, the coherence time and energy relaxation time of the higher energy levels exceed 20 $\mu$s \cite{Coherence-level1,Coherence-level3} and 30--140 $\mu$s \cite{relaxtion1,relaxtion2}, respectively. Different transformations between high levels can be achieved by applying consecutive $\pi$ pulses for each sequential transition frequency \cite{Coherence-level1}. The operation time of a two-qubit gate is 40 ns \cite{operation-time1,operation-time2} in the current superconducting systems. Therefore, our Fredkin gate operation is approximately 200 ns, which is within the coherence time of the multi-level state. The three-level system can also be encoded in long-lived ($\sim$ms coherence time) ground states of the diamond NV defect center $|m_s=\pm1\rangle$ and $|m_s=0\rangle$ \cite{NV2}. Alternatively, two computing states are encoded in the electron-spin states $|m_s=0\rangle$ and $|m_s=1\rangle$ ($\sim$ms coherence time), and the auxiliary state is encoded in the $^{13}$C ($^{14}$N) nuclear-spin $|m_I=+\frac{1}{2}\rangle$ or $|m_I=-\frac{1}{2}\rangle$ ($|m_I=0\rangle$ or $|m_I=-1\rangle$) with $\sim$s coherence time in the NV center, respectively \cite{NV3}. In the NV center, the single-qubit manuscript time is $\sim$10 ns for the electron spin and greater than 10 $\mu$s for the nuclear spin \cite{single-qubit}. Moreover, we can also encode computing states in the horizontal polarization state $|H\rangle$ and vertical polarization state $|V\rangle$, and the auxiliary state can be encoded in the spatial mode of a photon.




In summary, we have presented a general technique for synthesizing an $n$-control-qubit Fredkin gate with $2n+3$ CNOTs and $2n$ single-qubit operations. The synthesis of the three-qubit Fredkin gate is optimal in terms of the number of CNOTs, and all CNOTs are nearest-neighbor interaction CNOTs. Furthermore, using the available spatial DOF of the first target photon, we implemented three-qubit deterministic and non-deterministic optical Fredkin gates. The cost of the latter can be further reduced to 4 nearest-neighbor CNOTs, and the operation of this heralded gate is demonstrated through the use of a single photon detector. Our insights into Fredkin gate construction may contribute to simplification of large quantum circuits and even break through the theoretical lower bound of required CNOT-count.

\section*{Funding}

The work is supported by the National Natural Science Foundation of China under Grant No. 11604012, and the Fundamental Research Funds for the Central Universities under Grant Nos. FRF-TP-19-011A3 and 230201506500024, and a grant from China Scholarship Council.


\end{document}